\global\def\draftcontrol{0}

   \def\versionno{ n4resum}

\catcode`\@=11

\expandafter\ifx\csname draftcontrol\endcsname\relax\global\def\draftcontrol{0}
\fi

{\count255=\time\divide\count255 by 60
\xdef\hourmin{\number\count255}
\multiply\count255 by-60\advance\count255 by\time
\xdef\hourmin{\hourmin:\ifnum\count255<10 0\fi\the\count255}}
\def\draftdate{\number\month/\number\day/\number\year\ \ \ \hourmin }

\newcommand\makepapertitle{\par
  \begingroup
    \renewcommand\thefootnote{\@fnsymbol\c@footnote}%
    \def\@makefnmark{\rlap{\@textsuperscript{\normalfont\@thefnmark}}}%
    \long\def\@makefntext##1{\parindent 1em\noindent
            \hb@xt@1.8em{%
                \hss\@textsuperscript{\normalfont\@thefnmark}}##1}%
     \newpage
     \global\@topnum\z@   
     \@makepapertitle
     \thispagestyle{empty}\@thanks
  \endgroup
  \setcounter{footnote}{0}%
  \global\let\thanks\relax
  \global\let\makepapertitle\relax
  \global\let\@makepapertitle\relax
  \global\let\@thanks\@empty
  \global\let\@author\@empty
  \global\let\@date\@empty
  \global\let\@title\@empty
  \global\let\title\relax
  \global\let\author\relax
  \global\let\date\relax
  \global\let\and\relax
  \def\version{\let\version\@version\@gobble}
}
\def\@makepapertitle{%
  \newpage
   \ifnum\draftcontrol=1 {}
   \version\versionno
   \vskip 3em%
   \else
   \hfill\hbox to 3cm {\parbox{4cm}{\@pubnum}\hss}%
   \vskip 3em%
   \fi
   \begin{center}%
   \let \footnote \thanks
     {\LARGE {\@title}}%
     \vskip 1.5em%
     {\normalsize
       \lineskip .5em%
       \begin{tabular}[t]{c}%
         \@author
       \end{tabular}\par}%
     \vskip 1.5em%
     {\@bstract}%
     \end{center}%
     \vskip 1.5em
     \@date%
   \par
}

\gdef\@pubnum{}
\def\pubnum#1{%
  \gdef\@pubnum{#1}}

\gdef\@bstract{}
\def\Abstract#1{%
  \gdef\@bstract{%
   \parbox{\textwidth-0pc}{%
   \centerline{\bf Abstract}\penalty1000%
\kern.2cm%
\noindent
\renewcommand\baselinestretch{1.0}%
{#1}}}
}

\def\ps@paper{\let\@mkboth\@gobbletwo%
     \ifnum\draftcontrol=1
    \def\@oddfoot{\hbox to \textwidth{\tiny \versionno \hfil\tiny\draftdate}%
    \hskip -\textwidth \hbox to \textwidth{\hfil\rm\thepage\hfil}}%
     \else\def\@oddfoot{\hbox to \textwidth{\hfil\rm\thepage\hfil}}
     \fi
     \let\@evenfoot\@oddfoot
}

\def\body{\clearpage
          \pagestyle{paper}
    }

\def\@version#1{\ifnum\draftcontrol=1
\typeout{}\typeout{#1}\typeout{}
\vskip3mm\centerline{\hbox{\fbox{\normalsize{\tt DRAFT -- #1 -- }
                   {\draftdate}}}}\vskip3mm
\fi}
\let\version\@version
\long\def\eqlabel#1{\ifnum\draftcontrol=1
                    \tag@false  
                    \tag*{(\theequation) \hbox to -0.2cm{\hspace{0cm}\small{#1}\hss}}
                    \refstepcounter{equation}
                    \edef\@currentlabel{\theequation}
                    \ltx@label{#1}          
                    \else
                    \label{#1}
                    \fi
                    }
\let\st@bibitem\@bibitem
\let\st@lbibitem\@lbibitem
\ifnum\draftcontrol=1
  \def\@bibitem#1{%
    \st@bibitem{#1}\a@@label{#1}\ignorespaces}
  \def\@lbibitem[#1]#2{%
    \st@lbibitem[#1]{#2}\a@@label{#2}\ignorespaces}
  \def\a@@label#1{%
    \gdef\a@lab{\smash{\normalfont\small#1}}
    \ifvmode
      \if@inlabel
        \global\setbox\@labels\hbox{%
          \llap{\a@lab\let\a@lab\relax
                \kern\@totalleftmargin\kern\marginparsep}%
          \box\@labels}%
      \fi
    \fi}
\fi

\documentclass[12pt,letterpaper]{article}

\usepackage{amsmath,amssymb,array,calc,epsfig,rotating,bm}
\usepackage[sort]{cite}
\usepackage{graphicx}
\usepackage{psfrag,verbatim}
\usepackage{xcolor}

\ifnum\draftcontrol=1
\fi

\renewcommand\baselinestretch{1.25}
\renewcommand\section{\@startsection {section}{1}{\z@}%
                                   {-3.5ex \@plus -1ex \@minus -.2ex}%
                                   {2.3ex \@plus.2ex}%
                                   {\normalfont\large\bfseries}}
\renewcommand\subsection{\@startsection{subsection}{2}{\z@}%
                                   {-3.25ex\@plus -1ex \@minus -.2ex}%
                                   {1.5ex \@plus .2ex}%
                                   {\normalfont\normalsize\bfseries}}
\renewcommand\subsubsection{\@startsection{subsubsection}{3}{\z@}%
                                   {-3.25ex\@plus -1ex \@minus -.2ex}%
                                   {1.5ex \@plus .2ex}%
                                   {\normalfont\normalsize\it}}
\renewcommand\paragraph{\@startsection{paragraph}{4}{\z@}%
                                   {-3.25ex\@plus -1ex \@minus -.2ex}%
                                   {1.5ex \@plus .2ex}%
                                   {\normalfont\normalsize\bf}}

\numberwithin{equation}{section}

\def\revise#1       {\raisebox{-0em}{\rule{3pt}{1em}}%
                     \marginpar{\raisebox{.5em}{\vrule width3pt\
                     \vrule width0pt height 0pt depth0.5em
                     \hbox to 0cm{\hspace{0cm}{%
                     \parbox[t]{4em}{\raggedright\footnotesize{#1}}}\hss}}}}

\newcommand\nxt[1]  {\\\fnxt#1}
\newcommand{\ie}{{\it i.e.,}\ }
\newcommand{\eg}{{\it e.g.,}\ }

\def\calf         {{\cal F}}
\def\calg         {{\cal G}}

\def\caln         {{\cal N}}
\def\calo         {{\cal O}}

\def\del          {\partial}

\def\Re           {{\rm Re\hskip0.1em}}
\def\Im           {{\rm Im\hskip0.1em}}

\def\sqr#1#2{{\vcenter{\vbox{\hrule height.#2pt
 \hbox{\vrule width.#2pt height#1pt \kern#1pt
 \vrule width.#2pt}\hrule height.#2pt}}}}

\newcommand{\kk}{\mathfrak{q}}
\newcommand{\ww}{\mathfrak{w}}

\def\a{\alpha}

\def\w{\omega}

\def\g{\gamma}

\def\aa1{\phi}
\def\cc1{\psi}

\def\k{\kappa}

\def\l{\lambda}

\def\k{\kappa}

\def\t{\tau}
\def\s{\sigma}

\catcode`\@=12

\begin{document}


\title{\bf Sensitivity of holographic $\caln=4$ SYM plasma hydrodynamics to finite coupling corrections}

\date{July 13, 2018}

\author{
Alex Buchel\\[0.4cm]
\it Department of Applied Mathematics,\\
\it Department of Physics and Astronomy\\ 
\it University of Western Ontario\\
\it London, Ontario N6A 5B7, Canada;\\
\it Perimeter Institute for Theoretical Physics\\
\it Waterloo, Ontario N2J 2W9, Canada\\[0.4cm]
}

\Abstract{Gauge theory/string theory holographic correspondence for $\caln=4$
supersymmetric Yang-Mills theory is well under control in the planar
limit, and for large (infinitely large) 't Hooft coupling,
$\l\to \infty$. Certain aspects of the correspondence can be extended
including $\calo(\l^{-3/2})$ corrections. There are no reliable first
principle computations of the $\caln=4$ plasma non-equilibrium
properties beyond the stated order. We show extreme sensitivity of the
non-hydrodynamic spectra of holographic $\caln=4$ SYM plasma to
$\calo(\l^{-3})$ corrections, challenging  any  conclusions reached from
'resummation' of $\calo(\l^{-3/2})$ corrections. 
}

\makepapertitle

\body

\version\versionno
\tableofcontents

\section{Introduction and Summary}

The most studied example of the holographic correspondence relating gauge theories and
string theory is for the maximally supersymmetric $SU(N)$ $\caln=4$ Yang-Mills theory (SYM)
and type IIB string theory in $AdS_5\times S^5$ \cite{m1}. The number of colors $N$
of the SYM is related to the 5-form flux on the string theory side. Furthermore, the asymptotic
$AdS_5$ (or $S^5$) radius $L$ in units of the string length $\a'=\ell_s^2$ along with the
asymptotic value of the string coupling $g_s$ establishes a correspondence to the 't Hooft
coupling $\l$ on the SYM side:
\begin{equation}
\frac{L^4}{\a'^2}=4\pi g_s N=g_{YM}^2 N\equiv \l\,.
\eqlabel{larel}
\end{equation}
While  there has been tremendous progress over the years in developing the correspondence
\eg see \cite{Beisert:2010jr}, understanding the full parameter
space $\{N,\l\}$ is elusive. How much is exactly known depends on what questions
one asks. Thermal or non-equilibrium states of SYM plasma at strong coupling
are under control in the planar limit, $g_{YM}\to 0$ $N\to \infty$ with
$\l$ kept fixed, and (in addition) for large 't Hooft coupling $\l\gg 1$.
Only first subleading corrections $\propto \calo(\l^{-3/2})$ are computationally 
accessible \cite{Paulos:2008tn}. Here is a sample of SYM plasma results
including first subleading corrections in the limit $\l\to\infty$:
\nxt The thermal equilibrium free energy density of the SYM plasma is  
\cite{Gubser:1998nz,Pawelczyk:1998pb}
\begin{equation}
\calf=-\frac{\pi^2}{8}N^2 T^4 (1+15\g+\cdots) \,.
\eqlabel{f}
\end{equation}
\nxt The shear viscosity to the entropy density ratio is
\cite{Buchel:2004di,Buchel:2008ac,Buchel:2008sh}
\begin{equation}
\frac\eta s=\frac{1}{4\pi} (1+120\g+\cdots)\,.
\eqlabel{etas}
\end{equation}
\nxt The speed of the sound waves and the bulk viscosity is \cite{Benincasa:2005qc}
\begin{equation}
c_s^2=\frac 13+0\cdot \g+\cdots\,,\qquad \frac{\zeta}{s}=0\cdot \g+\cdots \,.
\eqlabel{csxi}
\end{equation}
\nxt A sample of the second-order transport coefficients (see \cite{Baier:2007ix,Grozdanov:2014kva}
for further details) is \cite{Buchel:2008bz,Buchel:2008kd}
\begin{equation}
\begin{split}
&\t_\Pi T=\frac{2-\ln 2}{2\pi}+\frac{375}{4\pi}\g+\cdots\,,\qquad \k=\frac{\eta}{\pi T}
(1-145 \g+\cdots)\,,\\
&\frac{\l_1 T}{\eta}=\frac{1}{2\pi}(1+215\g+\cdots)\,.
\end{split}
\eqlabel{secondorder}
\end{equation}
\nxt The plasma conductivity is
\cite{Waeber:2018bea}\footnote{The reference \cite{Waeber:2018bea}
corrects  the earlier computation \cite{Hassanain:2011fn}. }
\begin{equation}
\s=\s_{\infty}(1+125\g+\cdots)\,,
\eqlabel{sigma}
\end{equation}
where $\sigma_\infty$ is the plasma conductivity at infinite 't Hooft coupling.

In expressions \eqref{f}-\eqref{sigma} we introduced 
\begin{equation}
\g=\frac 18\zeta(3)(\a')^3\,.
\eqlabel{defg}
\end{equation}
Notice that as one proceeds from the corrections to the equilibrium quantities \eqref{f}
to the first-order \eqref{etas}, the second-order \eqref{secondorder} transport, the
conductivity \eqref{sigma}, the relative ``strength'' of the corrections grow.
The correction strength is even more dramatic, $\propto (10^4-10^5) \cdot \g$
to the spectra of the non-hydrodynamic plasma excitations
(the QNMs of the dual gravitational background) \cite{Stricker:2013lma,Steineder:2013ana}.
This observation led the authors of \cite{Waeber:2015oka} to propose the idea
of an effective resummation of $\g$-corrections. In a nutshell, on $\a'$-corrected
gravity side of the holographic correspondence one typically gets higher-derivative
bulk equations of motion. One can use the smallness of $\g$ to eliminated the higher-derivatives,
reducing the equations to the second-order ones, where $\g$ corrections affect the first order
derivatives at the most --- this is precisely what was done for example in computation of the shear viscosity in \cite{Buchel:2004di}. The next (new) step is to 'forget' that $\g$ must be small
in transformed equations, and instead treat the equations non-perturbatively in $\g$.
There are two effects of such a resummation at finite $\g$:
\begin{itemize}
\item it is possible to compute finite-$\g$ corrections to SYM observables
at infinitely large 't Hooft coupling;
\item one can discover new phenomena, which are absent in an infinite 't Hooft coupling limit.
\end{itemize}
It is the latter aspect of the resummation that should be subject to additional scrutiny in drawing
physical conclusions. In particular, following the resummation approach of \cite{Waeber:2015oka}, in \cite{Grozdanov:2016vgg} a new branch of the QNMs
was found --- these are (purported) SYM plasma excitations with $\Re(\ww)=0$. 
The physics of these new excitations was crucial to draw conclusions regarding properties of
$\caln=4$ spectral function at intermediate 't Hooft coupling \cite{Solana:2018pbk}.

To our knowledge, there is no discussion in the literature, even at a phenomenological level,
how robust is the resummation approach of \cite{Waeber:2015oka}? In this note we address this
question focusing on $\Re(\ww)=0$ branch of the QNMs identified in \cite{Grozdanov:2016vgg}. 
In the absence of the reliable corrections to type IIB supergravity we proceed as follows.
Recall the tree level type IIB low-energy effective action in ten dimensions 
taking into account the leading order string corrections  
\cite{Grisaru:vi,Gross:1986iv}
\begin{equation}
S =
 \frac{1}{2 \kappa_{10}^2 } \int d^{10} x \sqrt g
\ \bigg[ R - \frac{1}{2} (\partial \phi)^2 - \frac{1} {4 \cdot 5!} 
  (F_5)^2  +...+ 
\  \gamma \ e^{- {\frac32} \phi}  W + ...\bigg]   \  ,
\eqlabel{action10}
\end{equation}
where $W$ in a certain scheme 
is proportional to the fourth power of the Weyl tensor  
\begin{equation}
W =  C^{hmnk} C_{pmnq} C_{h}^{\ rsp} C^{q}_{\ rsk} 
 + \frac 12  C^{hkmn} C_{pqmn} C_h^{\ rsp} C^{q}_{\ rsk}\  . 
\eqlabel{w}
\end{equation}
A consistent (for the purpose of QNM spectra computation) Kaluza-Klein reduction of
\eqref{action10} on $S^5$ results in
\begin{equation}
S_5=
 \frac{1}{2 \kappa_{5}^2 } \int d^{5} x \sqrt g
\ \bigg( R +\frac{12}{L^2}+ 
\  \gamma   W \bigg)\,,
\eqlabel{action5}
\end{equation}
where $W$ is a five-dimensional equivalent of \eqref{w}.
We would like to stress that an effective action
\eqref{action5} includes all the terms at order $\g$
arising from string theory that are relevant
for physics of homogeneous and isotropic
thermal equilibrium states of $\caln=4$ SYM plasma,
and (non-)hydrodynamic fluctuations about them.
As it stands, results extracted from this action are valid only
up to $\calo(\g)$, \ie for infinitesimal $\g$, and thus
does not provide information about finite-$\g$ (finite 't Hooft
coupling) corrections to $\caln=4$ SYM observables.
The resummation procedure advocated in \cite{Waeber:2015oka}
follows the steps:
\begin{itemize}
\item{(a)} Derive relevant equations of motion from \eqref{action5}
to order $\calo(\g)$ inclusive.
\item{(b)} These equations contain higher
(than the second order) space-time derivatives. Using equations of motion
at order $\calo(\g^0)$, all the space-time derivatives (higher than the
first order) at order $\calo(\g)$  can be eliminated,
\eg see \cite{Buchel:2004di}. The resulting equations contain at most
second space-time derivatives and the space of perturbative in $\g$
solutions of these equations agrees (up to $\calo(\g)$) with the space
of solutions of perturbative equations in (a). 
\item (c) The proposal of \cite{Waeber:2015oka} is to treat
equations in $(b)$ as {\it exact} in $\g$. 
\end{itemize}

Clearly, there is no physical justification of step $(c)$ where one
extends, without  any modifications, EOMs valid at $\calo(\g)$ only. 
On can easily invent infinitely many resummation schemes
in the spirit of  \cite{Waeber:2015oka}. Here is one of them:
\begin{itemize}
\item{(A)} Derive relevant equations of motion from \eqref{action5}
to order $\calo(\g^k)$ inclusive, where $k\ge 1$ is an arbitrary 
integer.
\item{(B)} These equations contain higher
(than the second order) space-time derivatives. Using equations of motion
at orders $\calo(\g^m)$, $m< k$, all the space-time derivatives (higher than the
first order) at orders $\calo(\g^m)$, $1\le m\le k$   can be eliminated,
\eg see section 2. The resulting equations contain at most
second space-time derivatives and the space of perturbative in $\g$
solutions of these equations agrees (up to $\calo(\g^k)$) with the space
of solutions of perturbative equations in (A). 
\item (C) The {\it new} resummation is to treat
equations in $(B)$ as {\it exact} in $\g$. 
\end{itemize}

The {\it new} truncation and resummation procedure of $\g$-corrections
is as good (or as bad) as the one proposed in \cite{Waeber:2015oka}.
The purpose of our paper is precisely to test the robustness of the
different $k$ resummation schemes. Specifically,
we consider the simplest extension of the five-dimensional effective action
\eqref{action5}:
\begin{equation}
\tilde{S}_5=
 \frac{1}{2 \kappa_{5}^2 } \int d^{5} x \sqrt g
\ \bigg( R +\frac{12}{L^2}+ 
\  \gamma   W+\a\g^2 W^2 + \calo(\g^3)\bigg)\,,
\eqlabel{action5m}
\end{equation}
where we study a family of a constant $\a$ such that $|\a\g|\lesssim 1$.
Notice that the (phenomenological) action \eqref{action5m}
is assumed to be {\it exact}
up to order $\g^2$. At $\a=0$ the effective action \eqref{action5m}
is just $k=2$ representative of  the {\it new} resummation scheme explained
above. The order $\calo(\a)$ term
is one of the potential terms that could arise from real string theory computations
--- we do not claim that it is a dominant one (there could be other terms
at this order); neither do be know the precise value of $\a$.  
The purpose of introducing this $\a$-term is to illustrate that physical observables
does not necessarily have to be monotonic in $\g$. 
Given \eqref{action5m}, the  corrections at order
$\g^2$ arise from the second-order perturbation due to $\g W$ term, and directly due to the
first-order term in $\a$.  In the next section we present results of the computations.
In both cases,
\begin{itemize}
\item setting $\a=0$ but treating \eqref{action5} as \eqref{action5m};
\item fixing $\g=10^{-3}$ and exploring $|\a|\lesssim 100$, 
\end{itemize}
we find a dramatic variation in the spectrum of QNMs on the branch with $\Re(\ww)=0$.
Thus, we conclude that physics extracted from \eqref{action5} beyond the leading order in
$\g$ (in the absence of explicit and reliable computations of $\calo(\g^2)$ string theory corrections)
have to be treated with caution. We explicitly demonstrated this fact for some branches of the
spectra of QNMs, however this is also true for the relation between the black brane temperature
$T$ and the location of its horizon $r_0$ in the holographic dual to $\caln=4$ SYM plasma: 
from \eqref{temperature} the $\calo(\g^2)$ term (at $\a=0$) enters with
coefficient over 1400 larger than the $\calo(\g)$ term. While we believe that similar fate awaits
other observables,  the $\eta/s$ ratio in particular, this remains to be corroborated with explicit computations.
On a positive note, it is conceivable that some quantities in $\caln=4$ plasma exhibit $\calo(\g)$
features that remain qualitatively robust upon inclusion of higher order corrections.

\section{Technical details}

To facilitate comparison and readability, we follow notations of   \cite{Grozdanov:2016vgg}.

To order $\calo(\g^2)$, the black brane solution to the equations of motion following from
\eqref{action5m} is given by
\begin{equation}
ds^2=\frac{r_0^2}{u}\left(-f(u)Z_t\ dt^2+dx^2+dy^2+dz^2\right)+Z_u\ \frac{du^2}{4u^2 f}\,,
\eqlabel{5dmetric}
\end{equation}
where $f(u)=1-u^2$, $r_0$ is the parameter of non-extremality of the black brane geometry,
and
\begin{equation}
\begin{split}
&Z_t=1-15\g\ (5 u^2+5u^4-3 u^6)+\g^2\ \biggl(
\frac{161100}{7} u^{14} \a+\frac{30}{7} (-6630 \a+69720) u^{12}
\\&+\frac{36}{7} (-5525 \a-119560) u^{10}+\frac{45}{7}
(-4420 \a-11872) u^8+\frac{60}{7} (-3315 \a-7329) u^6\\&+\frac{90}{7}
(-2210 \a-6986) u^4+\frac{180}{7} (-1105 \a-3493) u^2
\biggr)\,,\\
&Z_u=1+15\g\ (5u^2+5u^4-19u^6)+\g^2\ \biggl(
(\frac{198900}{7} \a+89820) u^2+(\frac{198900}{7} \a+95445) u^4\\&
+(\frac{198900}{7} \a+20070) u^6
+(\frac{198900}{7} \a+57195) u^8+(\frac{198900}{7} \a+2744370) u^{10}\\&
+(\frac{198900}{7} \a-3680775) u^{12}-\frac{2321100}{7} u^{14} \alpha
\biggr)\,.
\end{split}
\eqlabel{ztzu}
\end{equation}
The $\g$-corrected Hawking temperature corresponding to the solution \eqref{5dmetric}
is
\begin{equation}
T=\frac{r_0}{\pi}\biggl(\ 1+15\ \g+\g^2\ \left(21420+\frac{47700}{7}\a\right)\ \biggr)\,.
\eqlabel{temperature}
\end{equation}

\begin{figure}[t]
\begin{center}
\psfrag{w}{{$-\Im(\ww^{[1]})$}}
\psfrag{l}{{$\ln(\ww^{[1]}_{\calo(\g^2)}/\ww^{[1]}_{\calo(\g)})$}}
\psfrag{g}{{$\g$}}
  \includegraphics[width=2.4in]{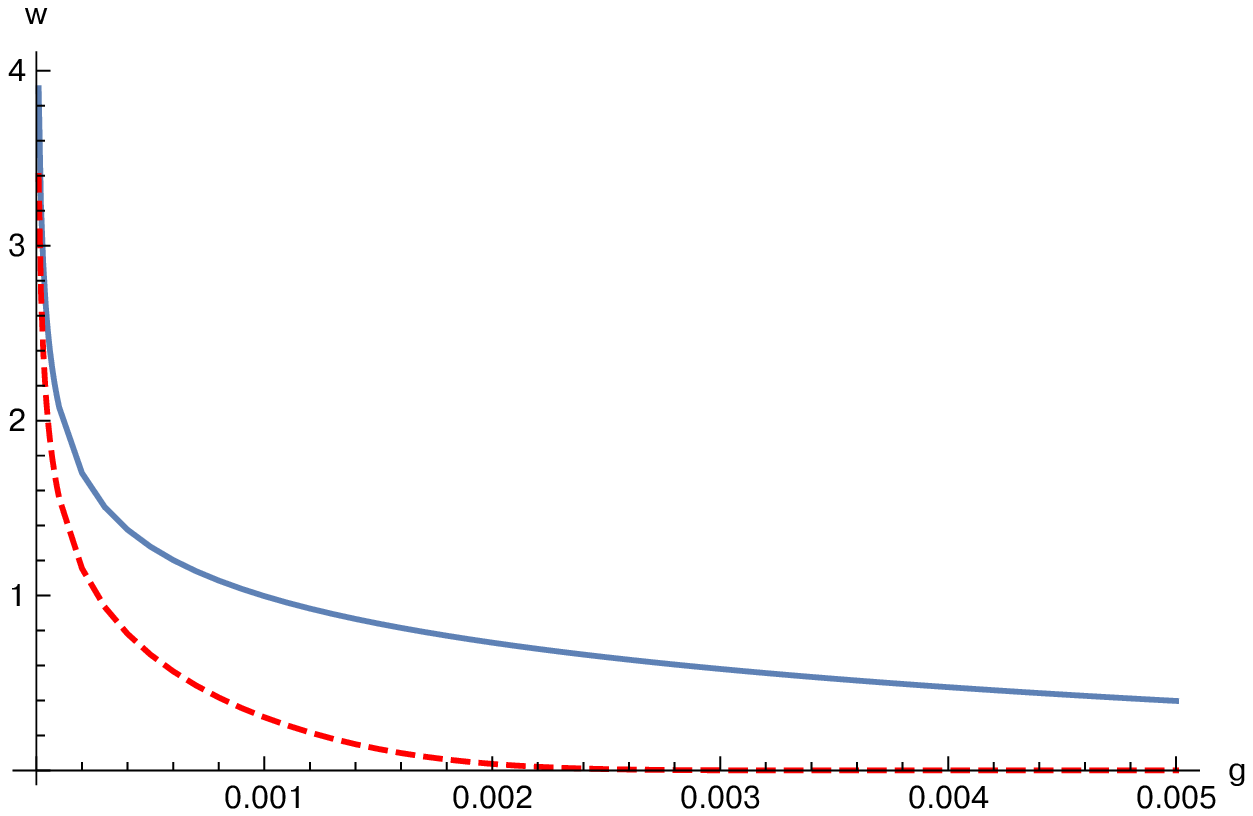}
  \includegraphics[width=2.4in]{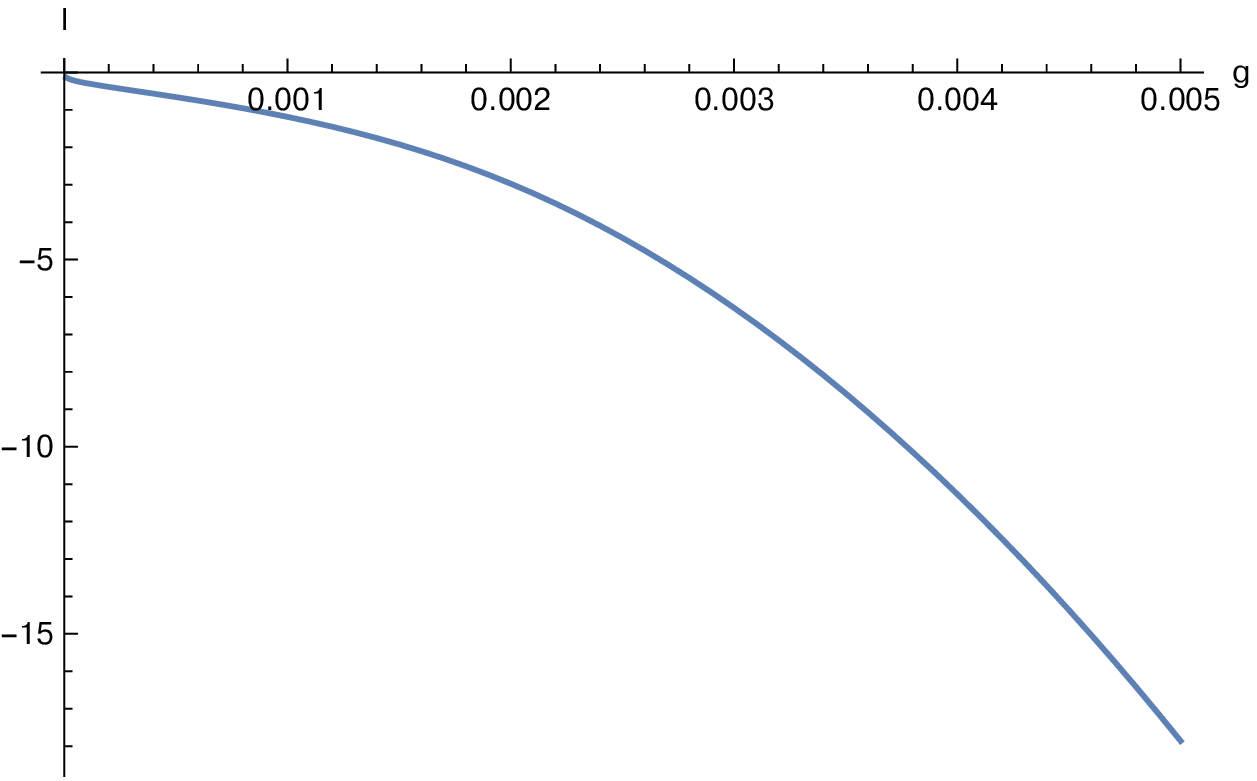}
\end{center}
 \caption{Left panel: the lowest QNM frequencies $\ww^{[1]}_{\calo(\g)}$ computed with \eqref{action5}
 (solid blue curve) and the lowest QNM frequencies  $\ww^{[1]}_{\calo(\g^2)}$ computed
 with \eqref{action5m} with $\a=0$ (dashed red curve).
 Right panel: log-comparison of the lowest QNM frequencies for different orders of the approximation
 of the gravitational effective action. 
}\label{figure1}
\end{figure}

\begin{figure}[t]
\begin{center}
\psfrag{w}{{$-\Im(\ww^{[1]})$}}
\psfrag{a}{{$\a$}}
  \includegraphics[width=4in]{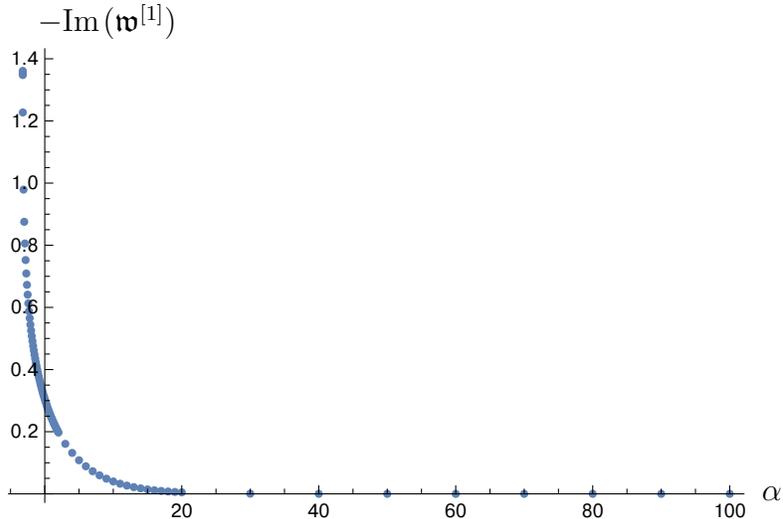}
\end{center}
 \caption{The lowest QNM frequencies $\ww^{[1]}_{\calo(\g^2)}$ computed
 with \eqref{action5m} at $\g=10^{-3}$ as a function of the phenomenological parameter
 $\a$.
}\label{figure2}
\end{figure}

{\bf Scalar channel} QNM equation takes form:
\begin{equation}
\del^2_uZ_1-\frac{1+u^2}{u(1-u^2)}\del_uZ_1+\frac{\ww^2-\kk^2(1-u^2)}{u(1-u^2)^2}Z_1
=\g\ \calg_1[Z_1]+\g^2\ \calg_{1,2}[Z_1]\,.
\eqlabel{z1}
\end{equation}
where $Z_1$ is a radial profile of the $h_{x}^{\ y}$ metric fluctuations.  
The explicit expression for $ \calg_1[Z_1,\del_u Z_1]$ can be found in \cite{Grozdanov:2016vgg},
and we compute
\begin{equation}
\begin{split}
&\calg_{1,2}=-\frac27 \biggl(
3144960 \a \kk^2 u^{11}+8052660 \a u^{12}-1075200 \kk^4 u^8+7878600 \a u^{10}\\
&+40025216 \kk^2 u^9
+75735891 u^{10}-994500 \a u^8-29659392 \kk^2 u^7+1741824 u^7 \ww^2\\
&+15490125 u^8-795600 \a u^6-40675194 u^6
-596700 \a u^4+604800 \kk^2 u^3-1040445 u^4\\
&-397800 \a u^2-843255 u^2-198900 \a-628740
\biggr) u\ \del_u Z_1+\frac{1}{7u (u^2-1)} \biggl(483840 \a \kk^4 u^{13}\\
&-17476020 \a \kk^2 u^{14}-258048 \kk^6 u^{10}+17945100 \a \kk^2 u^{12}-15661800 \a u^{12} \ww^2
\\
&+14363328 \kk^4 u^{11}-135086623 \kk^2 u^{12}-198900 \a \kk^2 u^{10}+2084400 \a u^{10} \ww^2-12425280 \kk^4 u^9
\\&+5246976 \kk^2 u^9 \ww^2+213413970 \kk^2 u^{10}-104522733 u^{10} \ww^2-198900 \a \kk^2 u^8\\
&+1686600 \a u^8 \ww^2
+100800 \kk^4 u^7-77651133 \kk^2 u^8+81113193 u^8 \ww^2-198900 \a \kk^2 u^6\\
&+1288800 \a u^6 \ww^2
+282240 \kk^4 u^5-1654800 \kk^2 u^6+3212370 u^6 \ww^2-198900 \a \kk^2 u^4\\
&+891000 \a u^4 \ww^2
+404775 \kk^2 u^4+1908900 u^4 \ww^2-198900 \a \kk^2 u^2+493200 \a u^2 \ww^2\\
&-644490 \kk^2 u^2
+1590435 u^2 \ww^2-95400 \a \kk^2+95400 \a \ww^2-301455 \kk^2+301455 \ww^2\biggr)\ Z_1
\,.
\end{split}
\eqlabel{g12}
\end{equation}
Note that the  EOM for $Z_1$ directly obtained from \eqref{action5m} involves terms $\propto \g$ or $\propto \g^2$
with (up to) forth-order derivatives in
$u$. Following \cite{Buchel:2004di}, higher-derivative "source'' terms with $\g$ dependence can be eliminated using
EOM at lower order. We implemented two different schemes:
\nxt all the higher-derivatives in $\g$-dependent source terms are eliminated using the $\calo(\g^0)$
EOM from \eqref{z1}:
\[
\del^2_uZ_1=\frac{1+u^2}{u(1-u^2)}\del_uZ_1-\frac{\ww^2-\kk^2(1-u^2)}{u(1-u^2)^2}Z_1\,;
\]
\nxt the functionals $\calg_1$ and $\calg_{1,2}$ (dependent on $Z_1$ and $\del_u Z_1$ only) are adjusted in such a way that
the perturbative solutions to \eqref{z1} agree with the perturbative solutions of the
higher-derivative order direct EOM for $Z_1$ to order $\calo(\g^2)$ inclusive.\\
The two reduction procedures are not equivalent: specifically,  $\calg_{1,2}$
differs\footnote{Nonetheless, we find that the QNM spectra computed within these two schemes
over the parameter range reported in figs.~\ref{figure1}-\ref{figure2} differ by less than $5\%$.}.  
Expression \eqref{g12} represents the result of the latter of the two reduction
schemes\footnote{I would like to thank the authors of \cite{Solana:2018pbk} for independent
confirmation of the technical details reported.}.

As in \cite{Grozdanov:2016vgg},
\begin{equation}
\ww=\frac{\w}{2\pi T}\,,\qquad \kk=\frac{q}{2\pi T}\,,
\eqlabel{wq}
\end{equation}
with the temperature given by \eqref{temperature}, and $\{\w,q\}$ begin the frequency and
the momentum of the non-hydro SYM plasma excitation.

We focus on QNMs with $\Re(\ww)=0$
at $\kk=0$. Thus, we need to solve (numerically) \eqref{z1} for $z_1$, defined as
\begin{equation}
Z_1=(1-u)^{-i\ww/2}u^2\ z_1(u)\,,
\eqlabel{defz}
\end{equation}
subject to a regular boundary conditions both as $u\to 0_+$ (the asymptotic $AdS_5$ boundary)
and $u\to 1_-$ (the black brane horizon):
\begin{equation}
\lim_{u\to 1_-} z_1=1\,,\qquad \lim_{u\to 0_+} z_1={\rm const}\ne 0\,.
\eqlabel{bc}
\end{equation}
Notice that \eqref{defz} automatically accounts for an incoming-wave boundary conditions for
$Z_1$ at the black brane horizon. Results of the numerical computations are presented in
figures \ref{figure1} and \ref{figure2}.
\nxt We confirm the computations of the QNM frequencies
determined in \cite{Grozdanov:2016vgg} and presented in Fig.~5 there.  
\nxt Left panel of fig.~\ref{figure1} presents the lowest  QNM frequencies
$\ww^{[1]}_{\calo(\g)}$ computed with \eqref{action5}
 (solid blue curve) and the lowest QNM frequencies  $\ww^{[1]}_{\calo(\g^2)}$ computed
 with \eqref{action5m} with $\a=0$ (dashed red curve) for a range of $\g\in[10^{-5}, 5\cdot 10^{-3}]$.
At $\g=10^{-5}$, the two approximations produce frequencies that differ by $\sim 13\%$.
As $\g$ increases, the difference becomes dramatic: at $\g=0.005$ the two frequencies differ
by a factor of $\sim 5\times 10^7$.
\nxt Fig.~\ref{figure2} presents results for  $\ww^{[1]}_{\calo(\g^2)}$ at
$\g=10^{-3}$ as parameter $\a$ varies within $[-3.21,100]$. The value of the
frequencies varies by a factor of $\sim 10^{11}$. Notice that the presented
QNM spectrum has a linear sensitivity to $\a$ about $\a=0$. This implies that,
lacking the precise knowledge of higher-derivative $\g$-corrections,
observables in $\caln=4$ SYM plasma does not have to be monotonic in
$\g$.

\section*{Acknowledgments}
I would like to thanks the organizers of HoloQuark2018 for an inspiring conference
in the culinary heart of Spain. I would like to acknowledge a  clear and interesting talk
by Jorge Casalderrey-Solana at the conference, which led to this note.
Research at Perimeter
Institute is supported by the Government of Canada through Industry
Canada and by the Province of Ontario through the Ministry of
Research \& Innovation. This work was further supported by
NSERC through the Discovery Grants program.


\end{document}